\documentclass[12pt,dvips]{article}
\usepackage{color}
\usepackage{bm}
\usepackage{url}
\usepackage{hyperref}
\usepackage{graphicx}
\usepackage{amsmath}
\usepackage{amssymb}
\usepackage{makeidx}
\title{Neutron spin structure with polarized deuterons \\[-.5ex]
and spectator proton tagging at EIC\footnote{Prepared for proceedings of 
Tensor Polarized Solid Target Workshop, Jefferson Lab, 
March 10-12, 2014}
\vspace{-1ex}}
\author{W.~Cosyn$^1$, V.~Guzey$^2$, D.~W.~Higinbotham$^3$, 
C.~Hyde$^4$, S.~Kuhn$^4$, \\[-.5ex]
P.~Nadel-Turonski$^3$, K.~Park$^4$, M.~Sargsian$^5$, 
M.~Strikman$^6$, C.~Weiss$^3$\footnote{Electronic address: 
\href{mailto:weiss@jlab.org}{weiss@jlab.org}}
\\[1ex]
\small $^1$ Ghent University, 9000 Gent, Belgium \\[-.7ex]
\small $^2$ Petersburg Nuclear Physics Institute, Gatchina, 
188300, Russia \\[-.7ex]
\small $^3$ Jefferson Lab, Newport News, VA 23606, USA \\[-.7ex]
\small $^4$ Old Dominion University, Norfolk, VA 23529, USA \\[-.7ex]
\small $^5$ Florida International University, Miami, FL 33199, USA \\[-.7ex]
\small $^6$ Pennsylvania State University, University Park, PA 16802, USA
\vspace{0ex} }
\date{September 18, 2014}
\begin{document}
\maketitle
\begin{abstract}
The neutron's deep--inelastic structure functions provide essential 
information for the flavor separation of the nucleon parton densities,
the nucleon spin decomposition, and precision studies of QCD phenomena 
in the flavor--singlet and nonsinglet sectors. 
Traditional inclusive measurements 
on nuclear targets are limited by dilution from scattering on protons,
Fermi motion and binding effects, 
final--state interactions, and nuclear shadowing at $x \ll 0.1$.  
An Electron--Ion Collider (EIC) would enable next--generation measurements 
of neutron structure with polarized deuteron beams and detection of 
forward--moving spectator protons over a wide range of recoil momenta
($0 < p_R < \textrm{several 100}\,\textrm{MeV}$ in the nucleus rest frame). 
The free neutron structure functions could be obtained by extrapolating the
measured recoil momentum distributions to the on-shell point.
The method eliminates nuclear modifications and can be applied to
polarized scattering, as well as to semi-inclusive and exclusive 
final states. We review the prospects for neutron structure measurements 
with spectator tagging at EIC, the status of R\&D efforts, 
and the accelerator and detector requirements.
\end{abstract}
\newpage
\section{Introduction}
The program of exploring short-range nucleon structure and strong 
interaction dynamics with high--energy lepton scattering relies
on measurements on the neutron as much as those on the proton target.
Neutron and proton data are needed to separate the isoscalar and isovector 
combinations of the deep-inelastic scattering (DIS) structure functions, 
which are subject to different short--distance dynamics 
(QCD evolution, higher--twist 
effects, small--$x$ behavior) and give access to different combinations 
of the parton densities (gluons and singlet quarks vs.\ non-singlet quarks). 
The unpolarized isovector structure function $F_{2p} - F_{2n}$ constrains 
the flavor composition of the nucleon's sea quark densities and permits 
study of small--$x$ dynamics in the non-singlet sector. 
In the polarized case the isovector structure function $g_{1p} - g_{1n}$ 
can be used to cleanly separate leading--twist and higher--twist dynamics,
taking advantage of its simple QCD evolution (no mixing with gluons).
The isoscalar combination $g_{1p} + g_{1n}$ 
can then be used to extract the polarized gluon density from the 
leading--twist QCD evolution. Both isospin combinations are needed to 
determine the flavor decomposition of the polarized quark densities
and their contributions to the nucleon spin.
Neutron and proton data combined are needed also to 
investigate the dynamical mechanism causing single--spin asymmetries 
in semi-inclusive DIS (where there are hints of surprisingly large 
isovector structures) and to constrain the 
generalized parton distributions in deeply--virtual Compton scattering.

Neutron structure is measured in high--energy scattering experiments with 
nuclear targets. The extraction of the free neutron structure functions 
from nuclear DIS data faces considerable challenges (see 
Ref.~\cite{Arneodo:1992wf} for a review). Nuclear binding modifies the
apparent neutron structure functions through the Fermi motion and
other dynamical effects. At moderate momentum transfers $Q^2 \sim
\textrm{few GeV}^2$ the nuclear cross sections are affected by 
final--state interactions involving other nucleons. At $x \ll 0.1$ the
large coherence length of the electromagnetic probe causes
quantum--mechanical interference of the amplitudes for scattering 
from different nucleons along its path, which results in shadowing 
and antishadowing effects in the cross section \cite{Frankfurt:2011cs}. 
In polarized measurements with light nuclei (deuteron
$D \equiv {}^2$H, ${}^3$He)
there is significant dilution from scattering on the proton(s).
One must know not only the degree of neutron polarization in the nucleus 
but also the spin dependence of the various nuclear modifications
of nucleon structure. Attempts to account for these nuclear effects 
theoretically are complicated by the fact 
that they arise as an average over different classes of configurations 
in the nuclear wave function, which cannot be separated in inclusive
measurements. It is clear that better experimental control of the nuclear 
environment is needed to improve the precision of neutron structure 
extraction. This is particularly relevant if a combination of neutron 
and proton data are to be used to study subtle QCD effects, such as 
the separation of leading and higher twist, non-singlet QCD evolution, 
and non-singlet small--$x$ behavior.

Inclusive deep--inelastic scattering from nuclei was measured in 
fixed--target experiments with electron and muon beams, at
electron--nucleon squared center--of--mass energies $s_{eN} \equiv s_{eA}/A$
in the range $s_{eN} \sim 10 - 900\, \textrm{GeV}^2$ (JLab 6 GeV, 
DESY HERMES, SLAC, CERN EMC/NMC, FNAL E665; 
see Refs.~\cite{Arneodo:1992wf,Malace:2014uea} 
for a review of the data). Measurements of neutron spin structure
with polarized nuclear targets were performed at SLAC, DESY HERMES, 
CERN SMC/COMPASS and 
JLab 6 GeV \cite{Kuhn:2008sy,Chen:2011zzp,Aidala:2012mv}. 
Both unpolarized and polarized nuclear measurements will be 
extended with the JLab 12 GeV Upgrade \cite{Malace:2014uea}. 
While these experiments have
provided basic information on neutron structure at $x \gtrsim 0.01$,
addressing the above questions will require data of much higher
precision and wider kinematic coverage.

The Electron--Ion Collider (EIC), proposed as a next-generation 
facility for nuclear physics, would dramatically expand the opportunities
for high-energy scattering on polarized light nuclei 
($D$, ${}^3$He, \ldots) and measurements of neutron 
structure.\footnote{For a general overview of the 
medium--energy EIC physics program, including measurements with 
proton beams, see e.g.\ Ref.~\cite{Accardi:2011mz}.}
The medium--energy EIC designs recently developed provide
electron--nucleon squared center--of--mass energies in the range 
$s_{eN} \sim 250-2500 \, \textrm{GeV}^2$ at 
luminosities up to $\sim 10^{34} \, \textrm{cm}^{-2} \, 
\textrm{s}^{-1}$ \cite{EIC-designs}. The wide kinematic coverage would 
permit definitive studies of the $Q^2$ evolution and small--$x$ behavior 
of structure functions; the high luminosity would enable measurements of
spin asymmetries and rare processes involving exceptional configurations.
Even more important, two specific capabilities provided by the EIC would allow
one to control the nuclear modifications and extract neutron structure
with unprecedented precision.

One capability are deuteron beams, especially polarized deuterons, 
as would be available for the first time with the JLab MEIC, thanks to the 
figure--8 shape of the ion ring designed to compensate the effect of spin 
precession \cite{Abeyratne:2012ah}. The deuteron is the simplest 
nucleus ($A = 2$); its wave 
function is known well up to large nucleon momenta $\sim$ 300 MeV, 
including the light-front wave function describing microscopic nuclear 
structure as probed in high-energy scattering 
processes \cite{Frankfurt:1981mk}. 
The deuteron has spin 1 and is mostly in the $L = 0$ configuration 
(S--wave), with a small admixture of $L = 2$ (D--wave), such that the 
proton and neutron are spin--polarized and their degree of polarization 
is known very well. Because there are only two nucleons the possibilities 
for final--state interactions are limited; in the configurations where they
can happen they can be estimated using theoretical models 
(see e.g.\ Refs.~\cite{Cosyn:2013uoa,Cosyn:2010ux} for a model in the 
resonance region $W \sim$ few GeV). 

The other capability is the detection of spectator nucleons emerging from the 
high--energy scattering process (``spectator tagging''). In collider 
experiments the spectator nucleons carry a fraction $\sim 1/A$ of the 
ion beam momentum and can be detected with appropriate forward detectors. 
The technique is uniquely suited to colliders: there is no target material 
absorbing low-momentum nucleons, and it can be used with polarized 
ion beams (longitudinal and transverse).
Spectator tagging is especially powerful in scattering on the
deuteron. It allows one to positively identify the active neutron
through proton tagging, and to control its quantum state through 
measurement of the recoil momentum.
Spectator proton tagging with unpolarized deuterium was explored in
a pioneering fixed-target experiment at JLab with 6 GeV beam energy
(CLAS BoNuS detector, covers recoil momenta $p_R \gtrsim 70\, \textrm{MeV}$) 
\cite{Tkachenko:2014byy} and will be studied further at 11~GeV.

In this note we summarize the potential of DIS on the deuteron with spectator 
proton tagging for precision measurements of the neutron structure 
functions at EIC. We describe the method for eliminating nuclear structure 
effects through on-shell extrapolation in the spectator proton momentum, 
present simulations of unpolarized and polarized observables \cite{LDRD}, 
and outline the accelerator and detector requirements. It should be 
noted that the various nuclear modifications of the nucleon's partonic 
structure are interesting physics topics in their own right (with connections 
to short-range $NN$ correlations, non-nucleonic degrees of freedom, etc.)
and can be studied with the same spectator tagging measurements
at larger recoil momenta \cite{Guzey:2014jva}.
\section{Neutron structure with spectator tagging} 
\begin{figure}
\parbox[c]{0.24\textwidth}{
\includegraphics[width=0.24\textwidth]{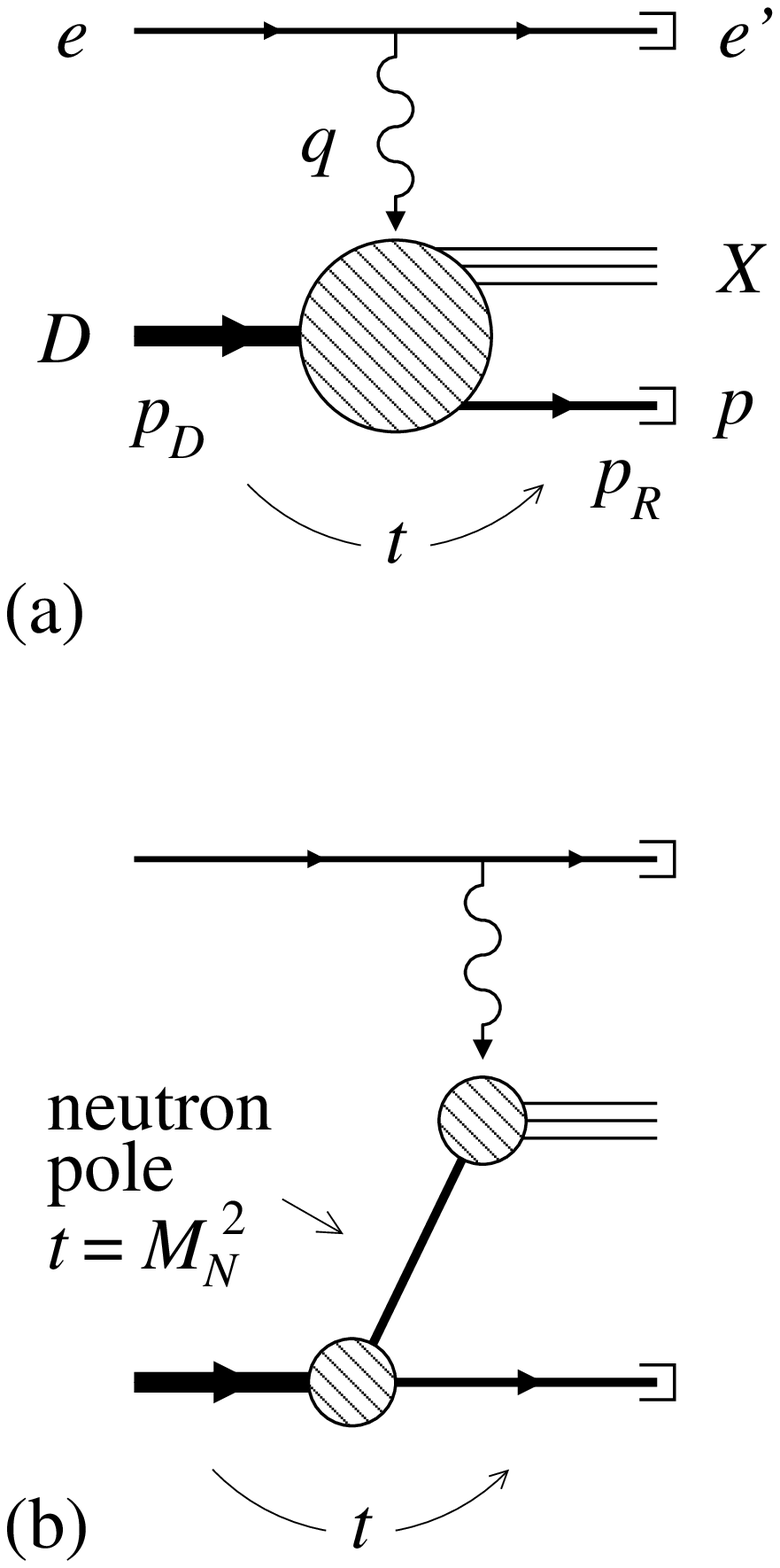}}
\hspace{0.1\textwidth} 
\parbox[c]{0.65\textwidth}{
\includegraphics[width=0.65\textwidth]{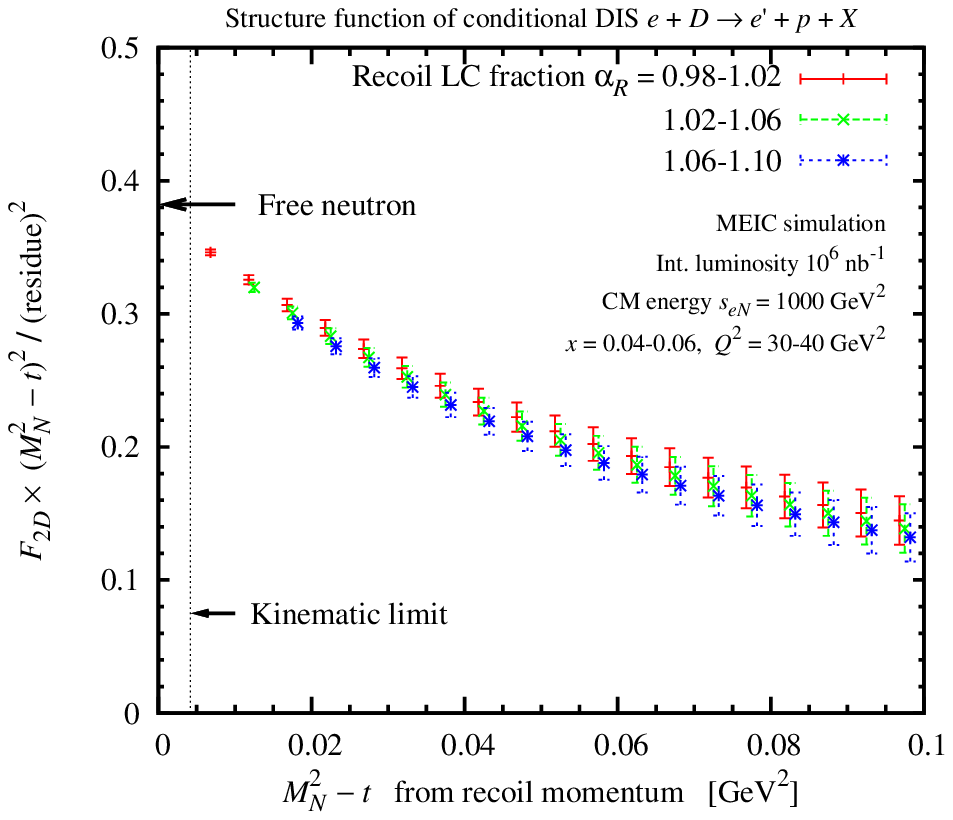}
\\[-3ex] {\small (c)}}
\vspace{1ex}
\caption[]{\small (a) Conditional DIS on the deuteron, 
$e + D \rightarrow e'+p+X$. 
(b) Neutron pole at $t = M_N^2$ arising from the impulse 
approximation diagram. (c) Simulated measurement of the
recoil momentum dependence with MEIC and on-shell extrapolation 
$t \rightarrow M_N^2$ \cite{LDRD}. The plot shows the conditional
structure function, with the pole factor $1/(M_N^2 - t)^2$ and
the residue removed, as a function of the off-shellness $M_N^2 - t$
calculated from the measured proton recoil momentum. 
The error bars indicate the expected statistical errors.
The three sets correspond to measurements in different intervals 
of the recoil light--cone momentum fraction $\alpha_R$. 
\label{fig:neutron}}
\end{figure}

The basic method for 
extracting the free neutron structure function with spectator tagging is 
described in Ref.~\cite{Sargsian:2005rm} (see Fig.~\ref{fig:neutron}).
One measures the cross section of conditional DIS 
$e + D \rightarrow e' + p + X$ as a function of the recoil proton momentum,
parametrized by the light--cone fraction $\alpha_R \equiv 
2 (E_R + p_R^z)/(E_D + p_D^z)$ and the transverse momentum $\bm{p}_{RT}$,
defined in a frame where the deuteron momentum $\bm{p}_D$ and 
the $\bm{q}$ vector are collinear and along the $z$--direction
(see Fig.~\ref{fig:neutron}a). 
A key variable is the invariant 4--momentum transfer between the
deuteron and the recoil proton, $t \equiv (p_R - p_D)^2$, calculated from
$\alpha_R$ and $\bm{p}_{RT}$. 
As a function of $t$ the scattering 
amplitude has a pole at $t = M_N^2$, which arises from the impulse 
approximation diagram of Fig.~\ref{fig:neutron}b and corresponds to
``neutron exchange'' in the $t$--channel. The residue at the pole is, 
up to a constant factor representing deuteron structure,
given by the structure function of the {\em free neutron,} evaluated at the
argument $\tilde x = x/(2 - \alpha_R)$, where $x = Q^2/(p_D q)$ is the
scaling variable with $0 < x < 2$. \footnote{The variables are defined such
that in the absence of nuclear binding $\alpha_R = 1$, and 
$x = \tilde x$ coincides with the usual scaling variable for scattering 
from a free nucleon.} It can be shown that nuclear 
binding and final--state interactions only affect the amplitude at
$M_N^2 - t > 0$, but not the residue at the pole \cite{Sargsian:2005rm}.
%
%

To extract the free neutron structure function one plots the measured
tagged cross section as a function of $t$, removes the pole factor 
$1/(M_N^2 - t)^2$, and extrapolates to 
$t \rightarrow M_N^2$ \cite{Sargsian:2005rm}. 
The pole in $t$ is extremely close to the physical region, 
so that the extrapolation can be performed with 
great accuracy; the minimum physical value of $M_N^2 - t$ is $\epsilon_D M_D$,
where $\epsilon_D$ is the deuteron binding energy.
The method is analogous to the Chew--Low extrapolation 
used to extract pion structure from $\pi N$ scattering data. 
Figure~\ref{fig:neutron}c shows a simulated on-shell extrapolation with 
MEIC pseudodata ($s_{eN} = 1000\, \textrm{GeV}^2$, integrated luminosity 
$10^{6}\, \textrm{nb}^{-1}$) \cite{LDRD}. One sees that the $t$--dependence
is very smooth. The extrapolation in $t$ is performed at {\em fixed} values of
the recoil light-cone fraction $\alpha_R$, such that the effective
$\tilde x$ in the neutron structure function remains constant along the way.
Figure~\ref{fig:neutron}c shows the pseudodata in three bins of 
$\alpha_R$ with average values 
$\langle \alpha_R \rangle = (1, \, 1.04, \, 1.08)$ 
and a width of $0.04$.
Note that the kinematic limit in $M_N^2 - t$ increases from $\epsilon_D M_D$
as $\alpha_R$ moves
away from unity. Comparison of the extrapolation results at different 
$\alpha_R$ allows one to confirm the universality of the nucleon pole
and provided an important test of the theoretical framework. Critical to 
the success of the method is the ability to measure the recoil proton
momentum with complete coverage down to $p_{RT} = 0$, and with a resolution 
$\Delta p_{\rm RT} \lesssim \textrm{20}\, \textrm{MeV}$
and $\Delta \alpha_R \lesssim 10^{-3}$.

The method described here permits a clean and model-independent extraction 
of free neutron structure and would enable precision measurements 
of $F_{2n}$ over the entire $x$--$Q^2$ range accessible with EIC.
Combined with proton data taken in the same kinematics it would 
determine the isovector structure function $F_{2p} - F_{2n}$, which 
constrains the flavor structure of the nucleon sea. The results
could be cross--checked against those from dilepton production in 
$pp/\bar pp$ collisions, which separate quark flavors through selection 
of $\gamma^\ast$ (Drell--Yan), $Z$, and $W^\pm$ events, and be used 
to test the universality of the sea quark densities \cite{Martin:2009iq}. 
The isovector $F_{2p} - F_{2n}$ would also permit quantitative studies 
of small--$x$ dynamics in the non-singlet sector, which is largely 
unexplored and raises many interesting questions (QCD structure of 
Reggeon exchange, non-singlet diffraction, etc.). In practice such
studies will likely be limited to the region $x > 0.01$, as the
difference between proton and neutron structure functions becomes
very small at low $x$ [$(F_{2p} - F_{2n})/(F_{2p} + F_{2n}) < 0.02$
at $x < 0.01$ for $Q^2 \sim \textrm{several 10 GeV}^2$] and would
be increasingly difficult to determine from separate measurements.
Lastly, the results for $F_{2n}$ itself would serve as stringent test 
for models of nuclear effects in heavier nuclei. 

DIS on the deuteron with spectator proton tagging could be used to address 
other interesting physics questions besides free neutron structure.
The recoil momentum dependence of the effective neutron structure function
in $e + D \rightarrow e' + p + X$ allows one to study the modification of 
the neutron's partonic structure as a 
function of its momentum in the deuteron wave function --- a qualitative
improvement compared to usual inclusive measurements. Theoretical
arguments suggest that the modification of the effective neutron
structure function in the deuteron is in first approximation 
proportional to the neutron's virtuality
$M_N^2 - t \approx 2 |\bm{p}_R|^2(\textrm{rest frame})$,
which is controlled by the measured recoil momentum \cite{Frankfurt:1985cv}.  
Furthermore, the modification is related in a simple way to the EMC 
effect in inclusive nuclear structure functions of heavier
nuclei ($A > 2$) and exhibits similar $x$--dependence.
Measurements of the recoil momentum dependence of the effective
neutron structure function in $e + D \rightarrow e' + p + X$ thus
provide direct insight into the dynamical origin of the EMC effect.
In particular, measurements at large recoil momenta 
$|\bm{p}_R|(\textrm{rest frame}) \gg 100\, 
\textrm{MeV}$ could verify a possible connection between the EMC 
effect and short--range $NN$ correlations in 
nuclei \cite{Weinstein:2010rt,Hen:2013oha}. 

Measurements of the recoil momentum dependence in 
$e + D \rightarrow e' + p + X$ at finite $|\bm{p}_{RT}|$ and 
$\alpha_R \neq 1$ also provide a unique method for studying nuclear
final--state interactions in DIS at $Q^2 \sim \textrm{few GeV}^2$. 
The results of such measurements could be used to refine phenomenological 
models of the final--state interactions \cite{Cosyn:2013uoa,Cosyn:2010ux},
which would improve the precision of neutron structure extraction 
from inclusive nuclear scattering data (including $A > 2$).

\section{Neutron spin structure} 
\begin{figure}[t]
\center{\includegraphics[width=0.65\textwidth]{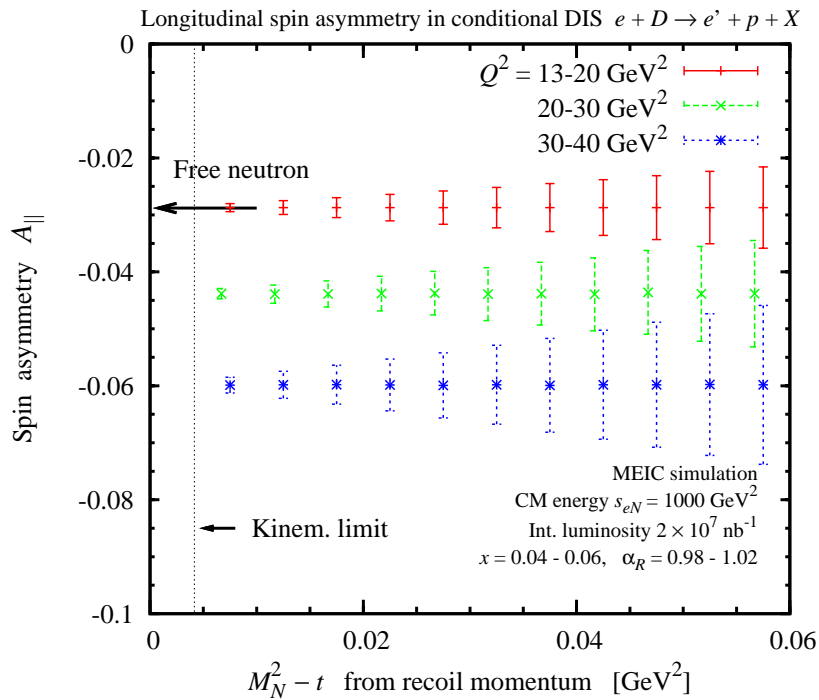}}
\caption{Simulated measurement of the longitudinal double spin
asymmetry $A_{\parallel}$, Eq.~(\ref{A_parallel}), in conditional
DIS on the deuteron with spectator proton tagging,
$\vec{e} + \vec{D} \rightarrow e' + p + X$ with MEIC.
The measured asymmetry is shown as a function of the off--shellness 
$M_N^2 - t$, calculated from the measured proton recoil momentum, 
in a single interval of the recoil light-cone momentum fraction $\alpha_R$.
The error bars indicate the expected statistical errors.
The three sets correspond to different values of $Q^2$. 
\label{fig:pol}
}
\end{figure}

Polarized DIS on the deuteron with spectator
proton tagging can be used to determine the free neutron spin structure 
function $g_{1n}$. The method for eliminating nuclear effects described 
above can straightforwardly be extended to double--polarized scattering. 
The simplest case is that of a longitudinally polarized electron beam 
colliding with a longitudinally polarized deuteron beam, where 
``longitudinal'' refers to the respective beam 
directions.\footnote{In the present MEIC design the electron and deuteron
beam collide with a finite crossing angle $\sim 50\, \textrm{mrad}$;
its effect on the net polarization along the $\bm{q}$ vector direction
can easily be calculated and is very small.} One measures the cross section 
of conditional double--polarized DIS $\vec{e} + \vec{D} \rightarrow e' + p + X$
as a function of the recoil proton momentum variables; the recoil proton 
polarization remains undetected and is summed over. The on-shell 
extrapolation in $t$ could in principle be performed directly with the 
polarized cross sections, using the same formulas as in the unpolarized case. 
It is more convenient, however, to work with the spin asymmetry
\begin{equation}
A_{\parallel} = \frac{\sigma(+ +) - \sigma(+ -)}{\sigma(+ +) + \sigma(+ -)} ,
\label{A_parallel}
\end{equation}
where $\sigma (\lambda_e, \lambda_D)$ denotes the conditional cross section 
for the spin projections $\lambda_e = \pm 1/2$ and $\lambda_D = \pm 1$. 
In Eq.~(\ref{A_parallel}) the nucleon pole $1/(M_N^2 - t)^2$ in the cross 
sections cancels between numerator and denominator (as do most of the
nuclear modifications at $M_N^2 - t > 0$), and the asymmetry naturally
has a smooth dependence on $t$. Figure~\ref{fig:pol} shows a 
simulated measurement of the conditional spin asymmetry at MEIC
as a function of $t$ ($s_{eN} = 1000\, \textrm{GeV}^2$, integrated luminosity 
$2 \times 10^{7}\, \textrm{nb}^{-1}$ = 20 times higher than in the
unpolarized simulation of Fig.~\ref{fig:neutron}c) \cite{LDRD}. 
The projections shown here were made using a simple model of deuteron 
structure that does not include the $D$--wave, assumes the same 
nuclear modification of polarized and unpolarized cross sections at 
$M_N^2 - t > 0$, and therefore does not have an explicit $t$--dependence 
of the asymmetry; in a more realistic model there would be a weak 
residual $t$--dependence at $M_N^2 - t > 0$ due to the said effects. 
Figure~\ref{fig:pol} shows that an accurate measurement of the 
conditional spin asymmetry could be made over a range of
$M_N^2 - t \lesssim 0.05 \, \textrm{GeV}^2$ with the proposed setup, 
and that the on-shell extrapolation could be performed as
in the unpolarized case. More elaborate simulations, assuming general 
deuteron polarization (mixed states, tensor polarization) and using
realistic deuteron wave functions, are in progress \cite{LDRD}. 
%
%

The on-shell (extrapolated) asymmetry can then directly be used to extract 
the {\em free neutron} spin structure function. An important simplification 
occurs because the $D$--wave contribution to the cross section is
very small at the on-shell point, as it is proportional to a higher power 
of the recoil proton rest--frame momentum. It implies that the deuteron wave 
function at the pole can be regarded as a practically pure $S$--wave, 
in which the neutron is completely polarized along the deuteron spin 
direction, and renders further analysis as simple as in the case of 
a polarized nucleon target:
\begin{equation}
A_{\parallel} \; (\textrm{on-shell}) \;\; \sim \;\;
D \, g_{1n}/F_{1n} 
\hspace{2em} \textrm{(up to terms $\sim M_N^2/Q^2$)},
\end{equation}
where $D$ is a kinematic factor (depolarization factor), and the neutron 
structure functions are evaluated at $\tilde x = x/(2 - \alpha_R)$
(complete expressions will be given elsewhere). Spectator tagging 
with on-shell extrapolation thus overcomes both the dilution and 
the neutron polarization uncertainties associated with inclusive 
nuclear measurements. 

The method outlined here would enable precision measurements of the neutron 
spin structure function $g_{1n}$ over a wide kinematic range without any 
model-dependent nuclear structure input. The neutron data thus obtained
could be used in several ways to advance our understanding of nucleon
spin structure and QCD. Combined with the proton data they would
determine the isovector polarized structure function $g_{1p} - g_{1n}$,
which exhibits particularly simple QCD evolution (no mixing with gluons) 
and can be used to cleanly separate leading--twist and higher--twist dynamics. 
They would also enable precision tests of the Bjorken sum rule, and possibly
a determination of the strong coupling $\alpha_s$ competitive with other
methods (see Ref.~\cite{Deur:2014vea} for a recent analysis using 
fixed--target data). 
The isoscalar structure function $g_{1p} + g_{1n}$ could be 
used to determine the polarized gluon density from the leading--twist 
QCD evolution. Proton and neutron data combined would allow one to
better separate quark and gluon contributions to the nucleon spin, 
and to determine the flavor composition of the quark spin. 
The status of global QCD fits to determine the polarized parton densities 
and the need for EIC data have have been discussed extensively in 
the literature. A dedicated study of the impact of spectator tagging
with polarized deuterons on polarized parton densities is planned.

%
%
\begin{figure}[t]
\center{\includegraphics[width=0.65\textwidth]{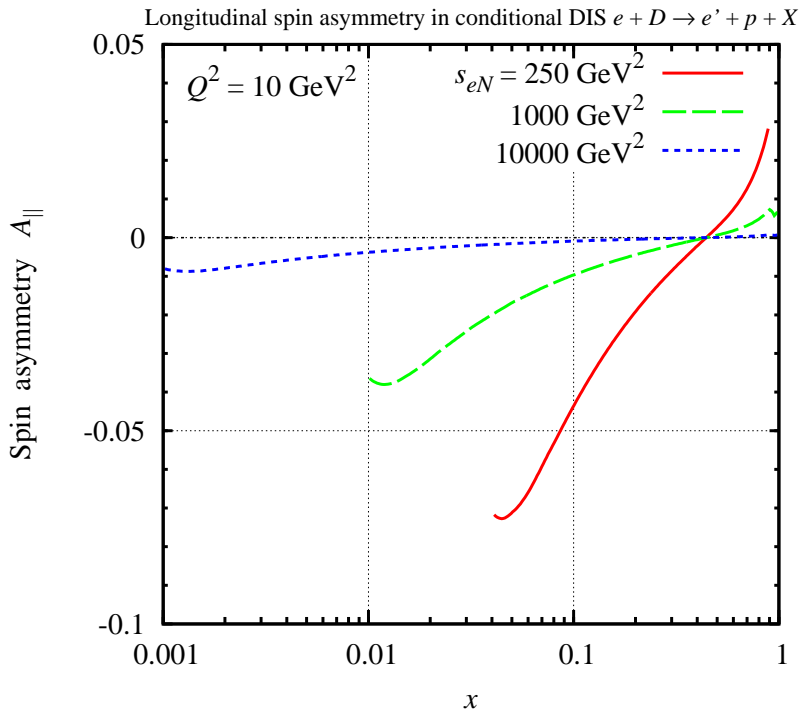}}
\caption{The observable double spin asymmetry 
$A_{\parallel}$, Eq.~(\ref{A_parallel}), in conditional
DIS on the deuteron with spectator proton tagging, 
$\vec{e} + \vec{D} \rightarrow e' + p + X$, as
a function of $x$, at $Q^2 = 10\, \textrm{GeV}^2$. 
The three curves correspond to measurements at squared 
electron--nucleon center-of-mass energies
$s_{eN} = 250, 1000$ and $10000 \, \textrm{GeV}^2$. 
The kinematic range in $x$ available at a given energy 
is $x > Q^2/s_{eN}$.
\label{fig:sdep}
}
\end{figure}
The center--of--mass energy of the electron--deuteron collision is an 
important parameter in the spin asymmetry measurements described here. 
According to Eq.~(\ref{A_parallel}) the magnitude of the observable 
spin asymmetry $A_\parallel$ is determined by the depolarization factor
\begin{equation}
D = \frac{y (2 - y)}{2 (1 - y)(1 + R) + y^2}
\hspace{2em} \textrm{(up to terms $\sim M_N^2/Q^2$)},
\end{equation}
where $R$ is the usual $L/T$ ratio of the unpolarized cross sections, and 
$y$ is the scaling variable measuring the electron fractional energy loss 
in the target rest frame. One sees that $D$ decreases proportionally to $y$ 
if $y \ll 1$. To maximize the observable $A_\parallel$ one wants
to keep $y$ at values of order unity. Now at fixed $x$ 
and $Q^2$ one has $y \approx Q^2/(x s_{eN})$, where $s_{eN} \equiv s_{eD}/2$. 
We conclude that one should do the measurement at squared CM energies 
$s_{eN}$ not much larger 
than $Q^2/x$, such that one can keep $y$ of order unity in the region
of interest. This is illustrated by Fig.~\ref{fig:sdep}, which shows
the projected observable spin asymmetry $A_\parallel$ at a fixed
$Q^2$ as a function of $x$, for different values
of $s_{eN}$. The projected values of the asymmetry here were calculated
using the parametrization of the neutron spin structure function 
provided by the global QCD fit of Ref.~\cite{Gluck:2000dy} 
(the differences to more recent fits are unimportant for the 
point illustrated here). One sees how the observable asymmetry at a given
$Q^2$ and $x$ decreases if measured at ``too large'' center--of--mass
energies. It underscores the need for an EIC design that can deliver
high luminosity over a \textit{range of center--of--mass energies}
appropriate to the measurements in question.
\section{Semi-inclusive and exclusive measurements}
The spectator tagging method described here can be extended to measurements
in which the final state of the DIS process on the neutron is analyzed 
further in coincidence with the spectator proton. Measurements of 
single--inclusive hadron production in the current fragmentation region
$e + D \rightarrow e' + p + h + X'$ (semi-inclusive DIS on the neutron)
could further constrain the partonic structure of the nucleon by tagging the 
charge and flavor of the struck quark in the neutron. Combined with proton
data such measurements could conclusively determine the flavor structure 
of the light nucleon sea (polarized and unpolarized) and separate 
strange and non-strange quark densities. Neutron measurements through
spectator tagging could also help in exploring the dynamical mechanisms 
generating single--spin asymmetries in semi-inclusive hadron production 
(there are hints of large isovector structures, possibly related to dynamical
chiral symmetry breaking) and determine the flavor structure of the
nucleon's intrinsic transverse momentum distributions. Finally, spectator
tagging could be used to measure exclusive processes on the neutron
(deeply--virtual Compton scattering, meson production), where the
isospin dependence is essential for verifying the reaction mechanism 
and determining the generalized parton distributions.
\section{Accelerator and detector requirements} 
The spectator tagging measurements 
described here require integrated forward detectors with complete 
coverage for protons with low recoil momenta relative to beam momentum 
per nucleon ($p_{RT} < 200 \, \textrm{MeV}, 
\; p_{R\parallel}/p(\textrm{beam}) \sim 0.8-1.2$), and
sufficient recoil momentum resolution ($\Delta p_{RT} 
\lesssim 20 \, \textrm{MeV}, \;
\Delta p_L/p_L \sim 10^{-4}$). The MEIC interaction region and forward 
detection system have been designed for this purpose and 
provide fully sufficient capabilities for the physics program outlined
here \cite{Abeyratne:2012ah}. The physics analysis of spectator tagging 
data also requires that the intrinsic momentum spread in the ion beam 
be sufficiently small to
allow for accurate reconstruction of the actual recoil momentum at
the interaction vertex. Simulations show that with the MEIC beam 
parameters the ``smearing'' of the kinematic variables is very moderate 
and does not substantially affect the physics analysis (the resulting
uncertainty in $t$ is of the order $\sim 0.005\, \textrm{GeV}^2$ --- 
the bin size in Fig.~\ref{fig:neutron}c) \cite{LDRD}.

The neutron spin structure measurements rely on the polarized deuteron 
beams that would become available with MEIC. In the figure--8 ring no 
significant loss of polarization is expected during acceleration and 
storage of either protons or deuterons, so that the polarization is
essentially maintained at the source level. It is expected that a 
longitudinal vector polarization in excess of 70\% could be 
achieved for a deuteron beam with $\sim 50$ GeV 
per nucleon \cite{Abeyratne:2012ah}, 
which would be sufficient for the measurements described here.

An R\&D program in under way at JLab to develop simulation tools
for spectator tagging with EIC (cross section models, event generators)
and demonstrate the feasibility of such measurements~\cite{LDRD}. 
The tools are being made available to users and can applied to 
a variety of processes of interest. Information about available
resources may be obtained from the authors.
\section*{Acknowledgments}
This material is based upon work supported by the U.S. Department of Energy, 
Office of Science, Office of Nuclear Physics under contract DE-AC05-06OR23177.
W.~Cosyn is supported by Research Foundation Flanders.
\\
Notice: Authored by Jefferson Science Associates, 
LLC under U.S.\ DOE Contract No.~DE-AC05-06OR23177. The U.S.\ Government 
retains a non--exclusive, paid--up, irrevocable, world--wide license to 
publish or reproduce this manuscript for U.S.\ Government purposes.
%
%

%
%
\end{document}